\documentclass[entropy,article,submit,moreauthors,pdftex,10pt,a4paper]{mdpi} 
\firstpage{1} 
\articlenumber{x}
\doinum{10.3390/------}
\pubvolume{xx}
\pubyear{2017}
\copyrightyear{2017}
\externaleditor{Academic Editor: name}
\history{Received: date; Accepted: date; Published: date}

\pdfoutput=1
\preto{\abstractkeywords}{\nolinenumbers}

\usepackage{url}

\Title{Conformity, Anticonformity and  Polarization of Opinions: Insights from a Mathematical Model of Opinion Dynamics}

\Author{Tyll Krueger $^{1}$, Janusz Szwabiński $^{2,}$* and Tomasz Weron $^{2}$}

\AuthorNames{Tyll Krueger, Janusz Szwabiński and Tomasz Weron}

\address{%
$^{1}$ \quad Department of Control Systems and Mechatronics, Wrocław University of Science and
Technology, Wrocław, Poland \\
$^{2}$ \quad Faculty of Pure and Applied Mathematics, Wrocław University of Science and
Technology, Wrocław, Poland}

\corres{Correspondence: janusz.szwabinski@pwr.edu.pl; Tel.: +48-71-320-3184}

\abstract{Understanding and quantifying polarization in social systems is important because of many reasons. It could for instance help to avoid segregation and conflicts in the society or to control polarized debates and predict their outcomes. In this paper we present a version of the $q$-voter model of opinion dynamics with two types of response to social influence: conformity (like in original $q$-voter model) and anticonformity. We put the model on a social network with the double-clique topology in order to check how the interplay between those responses impacts the opinion dynamics in a population divided into two antagonistic segments. The model is analyzed analytically, numerically and by means of Monte Carlo simulations. Our results show that the systems undergoes two bifurcations as the number of cross-links between cliques changes. Below the first critical point consensus in the entire system is possible. Thus two antagonistic cliques may share the same opinion only if they are loosely connected. Above that point the system ends up in a polarized state.}


\begin{document}



\section{Introduction}

\label{sec:introduction}

What do affirmative action and gun control~\cite{TAB:LOD:06}, same-sex marriage and sexual minority rights \cite{WOJ:PRI:10}, abortion~\cite{MOU:SOB:01}   stem cell research~\cite{BIN:DAL:BRO:SCH:09}, global warming~\cite{MCC:DUN:11}, attitudes toward political candidates~\cite{MEF:CHU:JOI:WAK:GAR:06} or the recent refugee crisis in Europe~\cite{SEE:14} have in common? All these keywords are examples of topics known to ignite polarized debates in the society. Studying them could thus shed more light on the phenomenon of polarization, which is one of the central issues in the recent opinion dynamics' research. Polarization is understood here as a situation in which a group of people is divided into two opposing parties having contrasting positions~\cite{DIM:EVA:BRY:96}. It is sometimes referred to as bi--polarization~\cite{MAS:FLA:13} to distinguish it from the so called group polarization, i.e. the tendency for a group to make decisions that are more extreme than the initial inclination of its members~\cite{ISE:86,SUN:02}.

Understanding and quantifying polarization is important because of many reasons. It could for instance help to avoid segregation and conflicts in the society~\cite{DIM:EVA:BRY:96} or to control polarized debates and predict their outcomes~\cite{WAL:91}. There are numerous theories~\cite{FRE:56,HAR:59,AXE:97,HEG:KRA:02,MAC:KIT:FLA:BEN:03} and many experimental attempts~\cite{IYE:HAH:09,STR:08,STR:10,KNO:MEN:11,GAR:09,GAR:CAR:LYN:11,GEN:SHA:11,MUT:MON:06,PRI:CAP:NIR:02} to explain the formation and dynamics of individuals' opinions, including mechanisms leading to polarization. As far as the theoretical part is concerned, mathematical and computational approaches are dominant in modelling of opinion dynamics~\cite{CAS:FOR:LOR:09}. In general, mathematical models allow for some theoretical \color{black} and/or numerical \color{black} analysis~\cite{GAN:LOW:STE:05,LOR:05,TOS:06,BRU:15,PAR:17,ALB:17}. They usually make some reasonable assumptions (e.g. a homogeneous and well-mixed population) to simplify the spreading process of opinions and focus at its representation at the macroscopic level. With the appearance of affordable high-performance computers simulation approaches to opinion dynamics are more and more popular~\cite{CAS:MUN:PAS:09,NYC:SZN:13,SZN:SZW:WER:14}. They usually build opinion formation models at individuals' level providing more detailed representation of realities at cost of higher computational complexity.

Agent-based models are one of the most powerful tools available for theorizing about opinion dynamics~\cite{LEI:14}. In many cases they act as in silico laboratories wherein diverse questions can be posed and investigated. There are already several attempts to apply such models to polarization. For instance, French~\cite{FRE:56}, Harary \cite{HAR:59} and Abelson~\cite{ABE:64} showed that consensus must arise in populations whose members are unilaterally connected unless the underlying social network is separated. According to Axelrod~\cite{AXE:97} local convergence may lead to global polarization. A number of papers has been devoted to explain polarization within the social balance theory, i.e. by accounting for sentiment in dyadic and/or triadic relations in social networks~\cite{MAR:KLE:KLE:STR:11,TRA:DOO:LEE:13}. In other computer experiments it has been shown that bridges between clusters in a social network (long-range ties) may foster cultural polarization if homophily and assimilation at the micro level are combined with some negative influence, e.g. xenophobia and differentiation~\cite{MAC:KIT:FLA:BEN:03, SAL:06}. Within the information accumulation systems the probability of reaching consensus has been found to decrease with the total number of interactions between agents that take place in the society~\cite{SHI:LOR:10}. From other studies it follows that polarization may be also induced by geometry of social ties~\cite{GAL:02}, mass media communication~\cite{MCK:SHE:06}  or some external actions of suitable controls (e.g. opinion leadership)~\cite{ALB:14,DUR:15}.   

Although the aforementioned models are very insightful, we still have some understanding gaps concerning polarization. One of the recent examples is the impact of new communication channels \color{black} like websites, blogs and social media \color{black}  on polarization. \color{black}
In particular social media services are by definition a space for information exchange and discussion. They shrink distances and facilitate communication among people of various backgrounds \color{black}. There are two competing hypotheses~\cite{LEE:CHO:KIM:KIM:14}. The first one states that people tend to expose themselves to likeminded points of view and rather avoid dissimilar perspectives. As a consequence, they form more extreme opinions in the direction of their original inclination~\cite{SUN:01,VAN:BRY:05}, which leads to both group and bi--polarization. Tools such as filtering and recommendation systems built in social media are considered to amplify this tendency. According to the other hypothesis new media enable people to encounter more diverse views and thus to have balanced opinions on different hot topics~\cite{BIM:08,PAP:02}. The empirical evaluation of these two hypotheses is inconclusive. Some studies have shown that people are more likely to select information sources consistent with their opinions or believes~\cite{IYE:HAH:09,STR:08,STR:10,KNO:MEN:11}. Cognitive dissonance, i.e. the mental stress or discomfort experienced by an individual holding two contradictory opinions at the same time, has been identified as one of the possible triggers of such behaviour~\cite{FES:57,KLA:60}. On the other hand there are some findings that individuals do not avoid information sources representing opposing points of view~\cite{GAR:09,GAR:CAR:LYN:11,GEN:SHA:11}. Some theories state that exposure to dissimilar views may have depolarizing effect, because it stimulates critical thinking~\cite{DEL:COO:JAC:04}. This effect has been observed in a series of experiments~\cite{MUT:MON:06,PRI:CAP:NIR:02}. 

Recently an Ising-type agent-based model of a social system has been presented to study if and how a combination of different responses to social influence may lead to polarization in a segmented network~\cite{SIE:SZW:WER:16}. \color{black} The model was based on the $q$-voter model of binary opinion dynamics~\cite{CAS:MUN:PAS:09} with an additional type of social response - anticonformity. From the statistical physics point of view the model fall into the category of quenched disorder models, which are known to be hard to analyze mathematically~\cite{LIU:05}. In this work we introduce an annealed version of that model that allows for mathematical treatment. We find a limiting dynamic system for a model of infinite size, which allows us to build the phase portrait of the model and gain some insight into its dynamics. The limiting system is solved numerically. However, we also calculate some of its characteristics analytically. Finally, we show with help of computer simulations that both the quenched and the annealed models of a finite size converge to the limiting system with the increasing number of agents.  \color{black}

The paper is organized as follows. In the next section we introduce the model and shortly describe how it differs from the one presented in~\cite{SIE:SZW:WER:16}. Then, we will investigate the model both analytically and numerically. Finally, some conclusions will be presented. 

\section{Materials and Methods}

\subsection{Basic assumptions}
\label{sec:basic}

We begin with a brief description of the assumptions of the model analysed in~\cite{SIE:SZW:WER:16}. We choose the $q$-voter model~\cite{CAS:MUN:PAS:09} as our modelling framework. Within the original model, $q$ randomly picked neighbours influence a voter to change his opinion. The voter conforms to their opinion if all $q$ neighbours agree. If they are not unanimous, the voter can still flip with a probability $\epsilon$. The unanimity rule 
is in line with a number of social experiments. For instance, it has been observed that a larger group with a nonunanimous majority is usually less efficient in terms of social influence than a small unanimous group~\cite{MYE:13,BON:05}. Moreover, Asch found that conformity \color{black}(i.e. matching attitudes, beliefs and behaviors to group
norms) \color{black} is reduced dramatically by the presence of a social supporter: targets of influence having a partner sharing the same opinion  were far more independent when opposed to a seven person majority than targets without a partner opposed to a three person majority~\cite{ASC:55}.

\begin{figure}[H]
\centering
\includegraphics[scale=0.4]{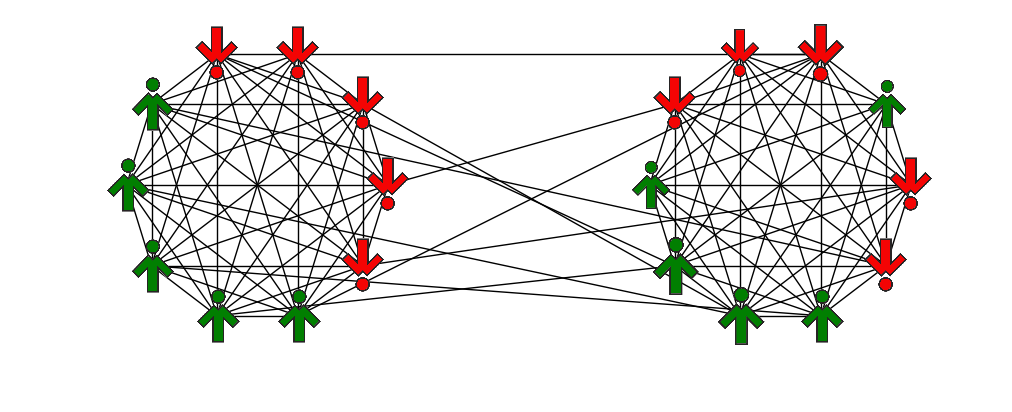}
\caption{An example of a double-clique network. The network consists of two separate complete graphs (cliques) with some cross links between them.\label{fig:two_clique}}
\end{figure}

From social networks analysis it follows that the existence of segments within a network may be correlated to polarization~\cite{CON:RAT:FRA:GON:FLA:MEN:11,NEW:06,ZAC:77}. We will thus assume that our social network is already modular. For the sake of simplicity we will put the model on the so called double-clique topology consisting of two complete graphs (cliques) connected by some cross links with each other~\cite{SOO:ANT:RED:08}. An example of such network is shown in Figure~\ref{fig:two_clique}. It should be emphasized that this is a rather strong assumption, because one actually cannot rule out the opposite possibility that segmentation is induced or intensified by polarization. However, analysis of the casual connection between the network segmentation and the polarization is beyond the scope of this work and will be addressed in a forthcoming paper.  

\begin{figure}[H]
\centering
\includegraphics[scale=0.3]{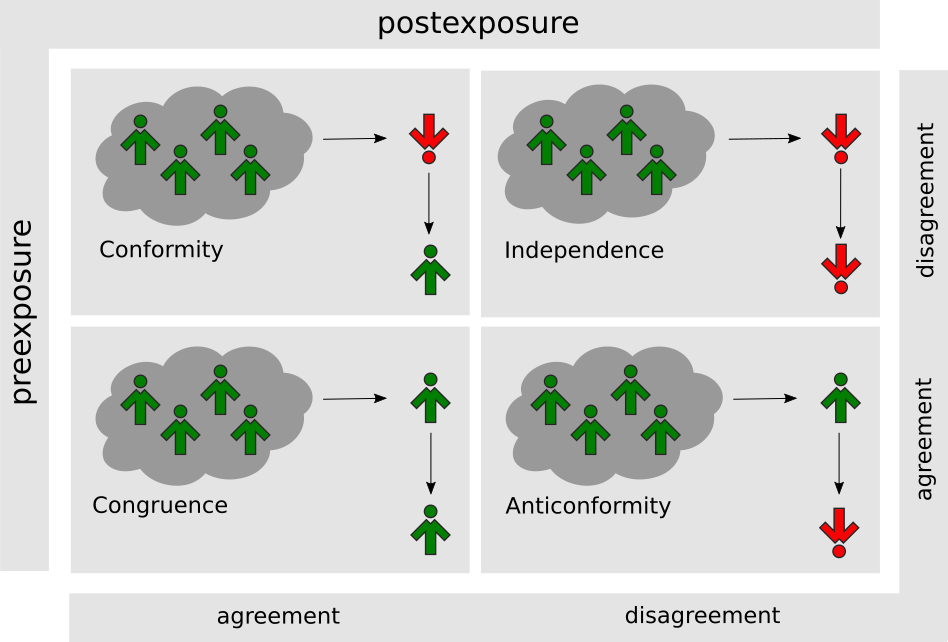}
\caption{Possible responses to the social influence according to the diamond model~\cite{WIL:63,NAI:86,NAI:DON:LEV:00,NAI:DOM:MAC:13}. Here presented within a $q$-voter model framework with $q=4$~\cite{CAS:MUN:PAS:09}. The source of influence is a group consisting of unanimous agents (schematically pictured as a cloud). The "up" and "down" spins (arrows) represent agents with two different opinions on a single issue. \label{fig:diamond}}
\end{figure}

The original $q$-voter model uses conformity, i.e. the act of matching attitudes and opinions to group norms, as the only response to social influence. Other possible types of responses has been described for instance within the diamond model~\cite{WIL:63,NAI:86,NAI:DON:LEV:00,NAI:DOM:MAC:13} and are shown in Figure~\ref{fig:diamond}. The anticonformity \color{black}(i.e. challenging the position or 
actions of a group) \color{black} representing negative ties is of particular interest, because from the social balance theories it follows that both positive and negative ties are needed for the polarization to emerge and to prevail~\cite{TRA:DOO:LEE:13}.Thus we will add this type of response to our model to \textbf{check how the interplay between conformity and anticonformity impacts its dynamics}.  

\color{black}
It is worth to note here that the anticonformity as a type of social response was introduced into binary models of opinion dynamics for the first time probably by Serge Galam, who used the notion of `contrarians' in order to describe agents always adopting opinions opposite to the prevailing choice of others~\cite{GAL:04}. In his seminal paper he showed that there exist a critical density of contrarians above which a system always ends up in a bi-polarized state.
\color{black}

We are however aware  that the assumption on negative influence is still a subject of an intense debate. There are several models able to explain polarization without any kind of negative influence, for example the argument-communication theory of bi-polarization~\cite{MAS:FLA:13} or the bounded-confidence model~\cite{HEG:KRA:02}. Some empirical studies on negative influence do not provide convincing support for this assumption as well~\cite{KRI:BAR:07}. 

There is an important point while speaking about anticonformity: it is relative. It turns out that in many settings multiple sources of norms are possible. As a consequence, conformity to one source can at the same time be anticonformity to another. For instance, a teenager's conformity to peers is very often anticonformity to his parents~\cite{NAI:DON:LEV:00}. Therefore we will assume within our model, that an agent strives for agreement within his own clique and simultaneously anticonforms to individuals from the other clique. 

Within the $q$-voter model all individuals are characterized by a single dichotomous variable, i.e. the model belongs to the class of binary models. At first glance this approach may seem unrealistic, because the opinions of individuals on specific subjects are expected to vary gradually. Therefore they should be rather represented by continuous variables~\cite{FRE:56,HAR:59,HEG:KRA:02,MAS:FLA:13}. But from empirical findings it follows that the distribution of opinions on important issues measured on some multivalued scale is often bimodal and peaked at extreme values~\cite{LEW:NOW:LAT:92,STO:GUT:SUC:LAZ:STA:CLA:50}. Moreover, many data on social networks are characterized by a semantic unicity, meaning that opinions and interactions of networks' members are restricted to a single domain or topic~\cite{GUE:MEI:CAR:KLE:13}. Very often those opinions may be interpreted as simple "yes"/"no", "in favour of"/"against" or "adopted"/"not adopted" answers~\cite{WAT:DOD:07}. In other words, in some situations the most important characteristics of the system under investigation may be  already captured by the relative simple models of binary opinions.

\subsection{The ``old'' (quenched disorder) model}

In this section we recall the model introduced in \cite{SIE:SZW:WER:16}. We consider a set of $2N$ agents, each of whom has an opinion on some issue that at any given time can take one of two values: $S_i=-1$  or $S_i=1$  for $i=1,2,\dots,2N$.  We will sometimes call these agents spinsons to reflect their dichotomous nature originating in spin models of statistical physics and humanly features and interpretation (\emph{spinson} = \emph{spin}+per\emph{son})~\cite{PRZ:SZN:WER:14,NYC:SZN:13}.

We put our agents on a double-clique network. It consists of two complete graphs of $N$ nodes connected with $L\times N^2$ cross links (Figure~\ref{fig:two_clique}). The parameter $L$ is the fraction of the existing cross links and $N^2$ -- their maximum number. The cross links between the cliques are chosen randomly and the resulting network does not evolve in time during simulation. Thus, from the statistical physics' point of view the model belongs to the class of models with quenched disorder~\cite{MAR:BON:MUN:14}. 

We will assume that a spinson strives for agreement within his own clique (conformity) and simultaneously challenges the opinions of individuals from the other clique (anticonformity). In other words we link the type of social response of agents with their group identity. To account for the possibility of acting as both conformist and anticonformist at the same time within the $q$-voter model, we introduce the notion of signals and slightly alter the concept of unanimity of the influence group. A signal is just a state of the neighbour when coming from spinson's clique or its inverted state otherwise. The target of influence changes its opinion  only if all members of the influence group emit the same signal. No other modification of the $q$-voter framework are needed to account for anticonformity. 

We use Monte Carlo simulation techniques with a random sequential updating scheme to investigate the model. Each Monte Carlo step in our simulations consists of $2\times N$ elementary events, each of which may be divided into the following substeps:
\begin{enumerate}[leftmargin=*,labelsep=4.9mm]
\item Pick a target spinson at random (uniformly from $2N$ nodes).
\item Build its influence group by randomly choosing $q$ neighboring agents.\label{influence group 1}
\item Convert the states of the neighbors into signals that may be received by the target. Assume that the signals of the neighbors from target's clique are equal to their states. Invert the states when from the other clique.
\item Calculate the total signal of the influence group by summing up individual signals of its members.
\item If the total signal is equal to $\pm q$ (i.e. all group members emit the same signal), the target changes its opinion accordingly. Otherwise nothing happens.
\end{enumerate}
Thus, our model is nothing but a modification of the $q$-voter with $\epsilon=0$ and an additional social response of spinsons. You may refer to~ \cite{SIE:SZW:WER:16} for further details of the model.\label{mc rules}

\subsection{New (``annealed'') version of the model}

In the model described in the previous section the cross links between the cliques are generated randomly at the beginning of a simulation and remain fixed while the system evolves in time. If the number of cross links is smaller than their maximum number $N^2$ (i.e. $L < 1$), some  agents may have no connections to the other clique, some others -- multiple ones. In other words, the agents may differ from each other because of the distribution of links between the cliques. While it can be handled with ease within a computer simulation, this feature constitutes usually a challenge for mathematical modelling \color{black} due to the necessity to perform a quenched average over the disorder~\cite{LIU:05} \color{black}. Thus, we decided to modify the model slightly for simpler mathematical treatment. 

Most of the assumptions presented earlier in this section hold, i.e. we consider $2 N$ agents living on a double-clique network. Each agent may be in one of the states $\left\{+1,-1\right\}$ representing its opinion on some issue. It seeks for agreement within his own clique (conformity) and simultaneously anticonforms to individuals from the other clique (anticonformity). And it changes its opinion if the members of an influence group emit the same signal. Just to recall,  a signal is just a state of the neighbour when coming from agents's clique or its inverted state otherwise.

If the size $N$ of each clique is large enough, then the number of links within a single clique is given by
\begin{equation}
\frac{(N-1)N}{2} \approx \frac{N^2}{2}.
\end{equation}
With the number of cross connections equal to $L\times N^2$ the quantity
\begin{equation}
p = \frac{LN^2}{LN^2+2\times\frac{N^2}{2}} = \frac{L}{1+L} 
\label{eq:p}
\end{equation}
gives us the probability of choosing one cross link out of all edges in the double-clique network. Assuming that every agent from one clique is connected with probability $p$ with an agent from the other clique, and with probability $1-p$ with an agent from its own clique we arrive at the new version of our model. Technically this approximation is nothing but an average of the original (quenched disorder) model over different configurations of cross links in the network. Thus it corresponds to annealed models from statistical physics~\cite{MAL:VAL:DAS:14}. 

\begin{figure}[H]
\centering
\includegraphics[scale=0.4]{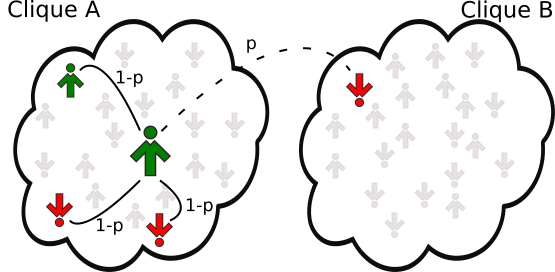}
\caption{Finding the influence group in the new version of the model. Each link of the target (the big green agent in the middle of clique A) is chosen independently. With probability $p$ it may point to an agent from the other clique (dashed line), with probability $1-p$ -- from the target's one (solid lines). See step~\ref{influence group 2} of the update procedure defined in paragraph~\ref{update} for more explanation. \label{fig:p}}
\end{figure}

The step~\ref{influence group 1} from the update rules  of the model defined in paragraph~\ref{mc rules} requires some adjustments:\label{update}  
\begin{enumerate}[leftmargin=*,labelsep=4.9mm]
\item Pick a target spinson at random (uniformly from $2N$ nodes).
\item Build its influence group by randomly choosing $q$ agents. \color{black} In the quenched disorder model we simply followed 4 randomly chosen links of the target to achieve that. Due to the setup of that model some targets usually had no cross connections, some others - multiple ones. Now the situation is different - each target has the same probability of being cross-connected and the actual links to other agents have to be built first. Thus, \color{black} for each member of the influence group we decide first which clique it will belong to (with probability $1-p$ to target's clique, with $p$ to the other one). Then we choose the member randomly from the appropriate clique (see Figure~\ref{fig:p}).\label{influence group 2}
\item Convert the states of the group members into signals.
\item Calculate the total signal of the influence group.
\item If the total signal is equal to $\pm q$ (i.e. all group members emit the same signal), the target changes its opinion accordingly. Otherwise nothing happens.
\end{enumerate}
It should be noted that a similar model, but with a broken symmetry between the cliques, was used in both the quenched and the annealed version to explain recurring fashion cycles~\cite{APR:KRU:MAR:SZN:16}. 

Let us denote the state of the $i$-th agent at the discrete time $\tau$ by $S_i(\tau)$. There are two natural quantities that fully describe the state of the system: the concentration of agents in state $+1$ and the average opinion (magnetization in physical systems)~\cite{NYC:SZN:13}. The concentration at time $\tau$ is defined as
\begin{equation}
c(\tau) = \frac{N^\uparrow(\tau)}{2N}~~\rightarrow ~~c(\tau)\in [0,1],
\end{equation}
where $N^\uparrow(\tau)$ is the number of agents in state $+1$ at time $\tau$. The average opinion is given by
\begin{equation}
m(\tau) = \frac{N^\uparrow (\tau)-N^\downarrow (\tau)}{2N}~~\rightarrow ~~m(\tau)\in [-1,1].
\end{equation}
Please note that there is a simple relation between these two quantities:
\begin{equation}
m(\tau) = 2c(\tau)-1.
\label{eq:m_c}
\end{equation}
Thus it actually does not matter which one will be chosen for representation of the state of the system. For the sake of convenience we will usually work with the concentration below. However, some of the results will be transformed to average opinions to allow for comparisons with the findings from~\cite{SIE:SZW:WER:16}.

\color{black} In order to easily detect polarized states in the system (i.e. all agents in state $+1$ in one clique and in state $-1$ in the other) we will often calculate the concentration separately for each clique:\color{black}
\begin{equation}
c_X(\tau) = \frac{N_X^\uparrow(\tau)}{N_X}~~~~,X=A,B,
\end{equation}
where $A$ and $B$ are the labels of the cliques.

The interpretation of $c_X$ is as follows:
\begin{itemize}[leftmargin=*,labelsep=5.8mm]
\item $c_X = 1$ -- positive consensus in clique $X$, i.e. all agents in that clique are in state $+1$,
\item $\frac{1}{2} < c_X < 1$ -- partial positive ordering in clique $X$, i.e. the majority of agents is in state $+1$,
\item $c_X = \frac{1}{2}$ -- no ordering in clique $X$, i.e. the numbers of agents in state $+1$ and $-1$ are equal,
\item $ 0 < c_X < \frac{1}{2}$ -- partial negative ordering in clique $X$, i.e. the majority of agents is in state $-1$,
\item $c_X = 0$ -- negative consensus in clique $X$, i.e. all agents in that clique are in state $-1$.
\end{itemize}

\subsubsection{Transition probabilities}

In each elementary time step the number of agents in state $+1$ in one clique -- say $A$ -- can increase by 1 only if:
\begin{enumerate}[leftmargin=*,labelsep=4.9mm]
\item a target from clique $A$ is chosen (probability $1/2$),
\item the target is in state $-1$ (probability $1-c_A$),
\item it flips, i.e. an influence group emitting signal $+q$ is chosen.
\end{enumerate} 
We can immediately write down the transition probability for such an event:
\begin{equation}
\Pr \left\{ N_{A}^\uparrow\left( t+\Delta _{N}\right) =N_{A}^\uparrow\left( t\right)
+1\right\} =\frac{1}{2}\left( 1-c_{A}(t) \right) \left[ \left(
1-p\right) c_{A}\left( t\right) +p\left( 1-c_{B}\left( t\right) \right)
\right]^{q}. \label{eq:transprob1}
\end{equation}
We have introduced here a scaled time $t=\frac{\tau}{2N}$ and a scaled time step $\Delta_N = \frac{1}{2N}$. We will use them below to derive a limiting dynamical system for the model. 
\begin{figure}[H]
\centering
\includegraphics[scale=0.4]{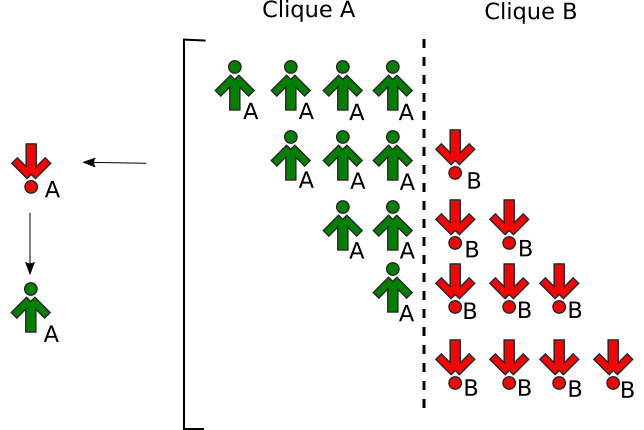}
\caption{Possible choices of the influence group in case $q=4$, that lead to an opinion flip of a spinson in clique $A$ being initially in state $S=-1$. See~\cite{SIE:SZW:WER:16} for further details. \label{fig:choices}}
\end{figure}
Moreover, one can easily check that the term of the form $(u+v)^q$ in the above equation is just the generating function for the probabilities of those compositions of $q$ members of an influence group which can cause an opinion switch event (see Figure~\ref{fig:choices}). 

Similarly, the number of spinsons in state $+1$ in clique $A$ decreases by 1 if:
\begin{enumerate}[leftmargin=*,labelsep=4.9mm]
\item a target from clique $A$ is chosen (probability $1/2$),
\item the target is in state $+1$ (probability $c_A$),
\item it flips, i.e. an influence group emitting signal $-q$ is chosen.
\end{enumerate} 
These conditions lead to the following transition probability:
\begin{equation}
\Pr \left\{ N_{A}^\uparrow\left( t+\Delta _{N}\right) =N_{A}^\uparrow\left( t\right)
-1\right\} = \frac{1}{2}c_{A}\left( t\right) \left[ \left( 1-p\right)
\left( 1-c_{A}\left( t\right) \right) +pc_{B}\left( t\right) \right]^{q}. \label{eq:transprob2}
\end{equation}

It is also possible that the number of agents in state $+1$ remains unchanged in elementary time step. The probability of this event is simply given by:
\begin{equation}
\Pr \left\{ N_{A}^\uparrow\left( t+\Delta _{N}\right) =N_{A}^\uparrow\left( t\right) \right\}
= 1-\Pr \left\{ N_{A}^\uparrow\left( t+\Delta _{N}\right) =N_{A}^\uparrow\left( t\right)
+1\right\} -\Pr \left\{ N_{A}^\uparrow\left( t+\Delta _{N}\right) =N_{A}^\uparrow\left(
t\right) -1\right\}. \label{eq:transprob3}
\end{equation}

After repeating analogous considerations for clique B we get:
\begin{eqnarray}
\Pr \left\{ N_{B}^\uparrow\left( t+\Delta _{N}\right) =N_{B}^\uparrow\left( t\right)
+1\right\} &=&\frac{1}{2}\left( 1-c_{B}\left( t\right) \right) \left[ \left(
1-p\right) c_{B}\left( t\right) +p\left( 1-c_{A}\left( t\right) \right)
\right] ^{q} \nonumber\\
\Pr \left\{ N_{B}^\uparrow\left( t+\Delta _{N}\right) =N_{B}^\uparrow\left( t\right)
-1\right\} &=&\frac{1}{2}c_{B}\left( t\right) \left[ \left( 1-p\right)
\left( 1-c_{B}\left( t\right) \right) +pc_{A}\left( t\right) \right] ^{q} \label{eq:transprob4}\\
\Pr \left\{ N_{B}^\uparrow\left( t+\Delta _{N}\right) =N_{B}^\uparrow\left( t\right) \right\}
&=&1-\Pr \left\{ N_{B}^\uparrow\left( t+\Delta _{N}\right) =N_{B}^\uparrow\left( t\right)
+1\right\} -\Pr \left\{ N_{B}^\uparrow\left( t+\Delta _{N}\right) =N_{B}^\uparrow\left(
t\right) -1\right\}\nonumber
\end{eqnarray}%
Thus, given the states of the cliques at time $t$, the expectations for the numbers of agents in state $+1$ at time $t+\Delta_N$ are given by the following expressions:
\begin{eqnarray}
E\left( N_{A}^\uparrow\left( t+\Delta _{N}\right) \right) &=& N_{A}^\uparrow\left( t\right) +%
\frac{1}{2}\left( 1-c_{A}\left( t\right) \right) \left[ \bar{p}c_{A}\left(
t\right) +p\left( 1-c_{B}\left( t\right) \right) \right] ^{q}\nonumber\\
& &-\frac{1}{2}%
c_{A}\left( t\right) \left[ \bar{p}\left( 1-c_{A}\left( t\right) \right)
+pc_{B}\left( t\right) \right] ^{q} \nonumber \\
E\left( N_{B}^\uparrow\left( t+\Delta _{N}\right) \right) &=&N_{B}^\uparrow\left( t\right) +%
\frac{1}{2}\left( 1-c_{B}\left( t\right) \right) \left[ \bar{p}c_{B}\left(
t\right) +p\left( 1-c_{A}\left( t\right) \right) \right] ^{q}\nonumber\\
& &-\frac{1}{2}%
c_{B}\left( t\right) \left[ \bar{p}\left( 1-c_{B}\left( t\right) \right)
+pc_{A}\left( t\right) \right] ^{q}\label{eq:expect}
\end{eqnarray}%
The abbreviation $\bar{p}=1-p$ was used in the above formulas.

\subsubsection{Asymptotic dynamical system}

We would like to derive from Eqs.~(\ref{eq:expect}) a limiting dynamical system for $N\rightarrow \infty$ in scaled time $t=\frac{\tau}{2N}$. Let us first divide the above equations by $N$:
\begin{eqnarray}
E\left( c_{A}\left( t+\Delta _{N}\right) \right) - c_{A}\left( t\right) &=&  
\Delta_N \left( 1-c_{A}\left( t\right) \right) \left[ \bar{p}c_{A}\left(
t\right) +p\left( 1-c_{B}\left( t\right) \right) \right] ^{q}\nonumber\\
& &-\Delta_N %
c_{A}\left( t\right) \left[ \bar{p}\left( 1-c_{A}\left( t\right) \right)
+pc_{B}\left( t\right) \right] ^{q} \nonumber \\
E\left( c_{B}\left( t+\Delta _{N}\right) \right) - c_{B}\left( t\right) &=& 
\Delta_N\left( 1-c_{B}\left( t\right) \right) \left[ \bar{p}c_{B}\left(
t\right) +p\left( 1-c_{A}\left( t\right) \right) \right] ^{q}\nonumber\\
& &-\Delta_N%
c_{B}\left( t\right) \left[ \bar{p}\left( 1-c_{B}\left( t\right) \right)
+pc_{A}\left( t\right) \right] ^{q}\label{eq:expect2}
\end{eqnarray}%
It is very likely that in the limit $N\rightarrow\infty$ the random variables $c_i = \frac{N^\uparrow_i}{N}$ localize and hence become almost surely equal to their expectations. We get
\begin{eqnarray}
\frac{c_{A}\left( t+\Delta _{N}\right) - c_{A}\left( t\right)}{\Delta_N} &=&  
\left( 1-c_{A}\left( t\right) \right) \left[ \bar{p}c_{A}\left(
t\right) +p\left( 1-c_{B}\left( t\right) \right) \right] ^{q}- %
c_{A}\left( t\right) \left[ \bar{p}\left( 1-c_{A}\left( t\right) \right)
+pc_{B}\left( t\right) \right] ^{q} \nonumber \\
\frac{c_{B}\left( t+\Delta _{N}\right) - c_{B}\left( t\right)}{\Delta_N} &=& 
\left( 1-c_{B}\left( t\right) \right) \left[ \bar{p}c_{B}\left(
t\right) +p\left( 1-c_{A}\left( t\right) \right) \right] ^{q} -c_{B}\left( t\right) \left[ \bar{p}\left( 1-c_{B}\left( t\right) \right)
+pc_{A}\left( t\right) \right] ^{q}\label{eq:expect3}
\end{eqnarray}
Taking the limit $N\rightarrow\infty$ and denoting the limiting variables $c_A$ and $c_B$ by $x$ and $y$ we arrive at
\begin{eqnarray}
x^{^{\prime }} &=&\left( 1-x\right) \left( \bar{p}x+p\left( 1-y\right)
\right) ^{q}-x\left( \bar{p}\left( 1-x\right) +py\right) ^{q}, \nonumber\\
y^{^{\prime }} &=&\left( 1-y\right) \left( \bar{p}y+p\left( 1-x\right)
\right) ^{q}-y\left( \bar{p}\left( 1-y\right) +px\right) ^{q}.\label{eq:dynamical}
\end{eqnarray}

\subsubsection{Annealed model as a birth-death process}

According to Eqs.~(\ref{eq:transprob1})-(\ref{eq:transprob4}) we have only two types of state transitions in each clique: `births', which increase the state variable by one, i.e. $N_X^{\uparrow}\rightarrow N_X^{\uparrow}+1$ ($X=A,B$), and `deaths',  which decrease it by one,  $N_X^{\uparrow}\rightarrow N_X^{\uparrow}-1$. Thus our model may be seen as two coupled birth-death processes~\cite{KNE:WEB:KRU:FRE:15}. Since such a process is relatively easy to simulate, we will use it as an additional benchmark while comparing the results for the quenched disorder and annealed models.

\section{Results}

All results presented in this section were obtained via symbolic and numerical calculations by making use of the Python's scientific stack~\cite{SCI:16}. Python codes needed to reproduce some of them may be found in the supplementary materials.

\color{black}
Although we will often use the case $q=3$ for presenting the results, there is no particular reason for choosing this value. We considered in our analysis influence groups of sizes ranging from 1 to 6. The upper bound of the group size was motivated by the conformity experiments by Asch~\cite{ASC:55}. Qualitatively, the results turned out to be independent on the actual value of $q$. However, with increasing $q$ the critical points were shifted towards higher values of the interaction parameter $p$ (see Fig.~\ref{fig:comparison}).
\color{black}

\subsection{Direction fields and stationary points}
\label{phase} 

\begin{figure}[H]
\centering
\includegraphics[scale=0.4]{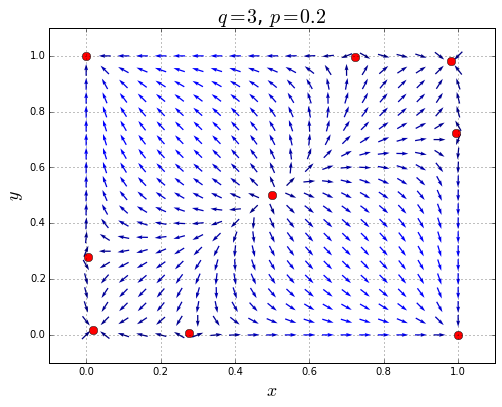}
\includegraphics[scale=0.4]{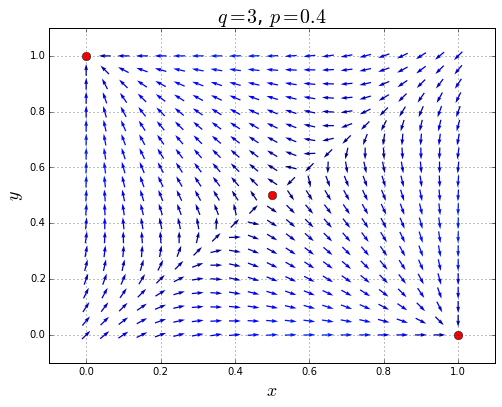}
\caption{Direction field of the dynamical system~(\ref{eq:dynamical2}) for $q=3$ and two different values of the interaction parameter $p$: $p=0.2$ (left plot) and $p=0.4$ (right plot). Red dots indicate stationary points of the system. \color{black}The parameters $q$ and $p$ are the size of the influence group and the probability of being connected to an agent from other clique, respectively.\color{black}\label{fig:quiver}}
\end{figure}

Our goal is to investigate the dynamical system given by Eq.~(\ref{eq:dynamical}),
\begin{eqnarray}
x^{{\prime }} &=&\left( 1-x\right) \left( \bar{p}x+p\left( 1-y\right)
\right) ^{q}-x\left( \bar{p}\left( 1-x\right) +py\right) ^{q}, \nonumber\\
y^{{\prime }} &=&\left( 1-y\right) \left( \bar{p}y+p\left( 1-x\right)
\right) ^{q}-y\left( \bar{p}\left( 1-y\right) +px\right) ^{q}.\label{eq:dynamical2}
\end{eqnarray}
It is customary to start such an analysis by plotting direction fields in the state space of the system~\cite{STR:94}. Examples of the fields for $q=3$ and two values of the parameter $p$ are shown in Figure~\ref{fig:quiver}. As a remainder: the solution trajectory through a given initial state is a curve in the state space which at every point is tangent to the field at that point. 
Several things are immediately clear from the picture shown in the figure. At $p=0.2$ the flows in the state plane suggest that there are nine stationary points (already marked with red dots). Some of these points are easy to classify. For instance, there are two attractors \color{black}(i.e. points toward which the system tends to evolve for a wide variety of initial conditions)  \color{black} at $(0,1)$ and $(1,0)$ corresponding to a polarized state of the system, i.e. all agents in one clique are in state $+1$ and in the other -- in state $-1$. Moreover, there are two other symmetric attractors close to the coordinates $(0,0)$ and $(1,1)$. It seems that the (almost) complete consensus is possible in our system 
for some initial configurations, at least for that particular value of $p$. The point $(0.5,0.5)$ is a repeller \color{black}(the system tends to evolve away from it) \color{black} and the remaining four points seem to be hyperbolic \color{black}(near such points the orbits of a two-dimensional, non-dissipative system resemble hyperbolas)\color{black}.

At $p=0.4$ (right plot in Figure~\ref{fig:quiver}) all hyperbolic points and the symmetric attractors disappear. The point $(0.5,0.5)$ becomes hiperbolic. The only remaining attractors are $(0,1)$ and $(1,0)$. Hence for higher values of $p$ the polarization of the system is the only possible asymptotic state. It should be noted that this findings recapture the results from~\cite{SIE:SZW:WER:16}.

To find the exact coordinates of the system we just set $x^{{\prime }}$ and $y^{{\prime }}$ equal to zero in Eq.~(\ref{eq:dynamical2}) and solve the resulting set of equations with respect to $x$ and $y$,
\begin{eqnarray}
0 &=&\left( 1-x\right) \left( \bar{p}x+p\left( 1-y\right)
\right) ^{q}-x\left( \bar{p}\left( 1-x\right) +py\right) ^{q}, \nonumber\\
0 &=&\left( 1-y\right) \left( \bar{p}y+p\left( 1-x\right)
\right) ^{q}-y\left( \bar{p}\left( 1-y\right) +px\right) ^{q}.\label{eq:dynamical3}
\end{eqnarray}

For $p=0.2$ and $q=3$ we get:
\begin{eqnarray}
&& P_1 = (0,1),~~ P_2 = (1,0) \nonumber\\
&& C_1 = (0.019, 0.019),~~ C_2 = (0.981, 0.981) \label{eq:fixed}\\
&& R_1 = (0.5,0.5)  \nonumber\\
&& U_1 = (0.005, 0.277),~~ U_2 =  (0.277, 0.005)\nonumber\\
&& U_3 = (0.722, 0.995),~~ U_4 = (0.995, 0.722)\nonumber
\end{eqnarray}
The linear stability analysis of these points reveal that indeed $P_1$, $P_2$, $C_1$ and $C_2$ are stable equilibria, $R$ is a repeller and the remaining points are hyperbolic ones, in agreement with our analysis of the direction field in Figure~\ref{fig:quiver}.

Similarly, for $p=0.4$ and $q=3$ we have:
\begin{eqnarray}
&& P_1 = (0,1),~~ P_2 = (1,0) \label{eq:fixed2}\\
&& R_1 = (0.5,0.5)  \nonumber\\
\end{eqnarray}
As before, $P_1$ and $P_2$ are stable, but $R_1$ is now hyperbolic. The remaining points disappeared (they became complex). 

\begin{figure}[H]
\centering
\includegraphics[scale=0.4]{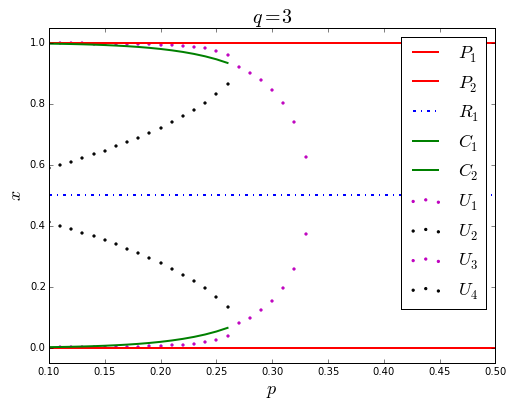}
\caption{Bifurcation diagram of the system given by Eq.~(\ref{eq:dynamical2}) for $q=3$. Solid lines indicate stable fixed points, the remaining ones - repellers and hyperbolic equilibria. At small values of $p$ the system has four attractors: $P_1$ and $P_2$ correspond to asymptotic polarization, $C_1$ and $C_2$ - to consensus. There is also a repeller $R_1$ and four hyperbolic points $U_1$--$U_4$. As $p$ increases a bifurcation happens at $p^*\simeq 0.26$. The four hyperbolic non-symmetric points $U_1$--$U_4$ disappear and the consensus equilibria $C_1$ and $C_2$ become hyperbolic. With further increase of $p$ the symmetric points eventually disappear and the repeller $R_1$ becomes hyperbolic. \label{fig:bif}}
\end{figure}

If we repeat the above calculations for other values of the parameter $p$ and plot the results, we get a bifurcation diagram showing how the dynamics of the system changes with $p$, i.e. with increasing degree of anticonformity in the system. The plot of the $x$ coordinates of the fixed points as functions of $p$ is shown in Figure~\ref{fig:bif}. The picture for the $y$ coordinates would look the same (but with rearranged labels of the unstable points $U_i$).    
Stable equilibria are indicated with solid lines. We see that the system has two attractors $P_1 = (0,1)$ and $P_2 = (1,0)$ as well as the unstable point $R_1 = (0.5,0.5)$ as stationary solutions independently of $p$. However, 
the system undergoes two phase transitions. At $p_1^\star\simeq 0.26$ the hyperbolic points $U_1$--$U_4$ disappear and the nontrivial symmetric fixed points $C_1$ and $C_2$ become hyperbolic. At $p_2^\star \simeq 0.32 $ these symmetric points disappear as well.

The above results were obtained numerically. However, in the special case of $q=2$ one can find both critical values of $p$ analytically. For a general value of $q$ an analytical computation of the second critical point is possible as well. We present corresponding calculations in Appendices~B and~C.

\subsection{Time evolution of the asymptotic system}

\begin{figure}[H]
\centering
\includegraphics[scale=0.5]{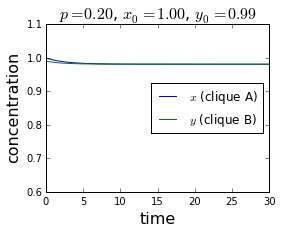} \  \includegraphics[scale=0.5]{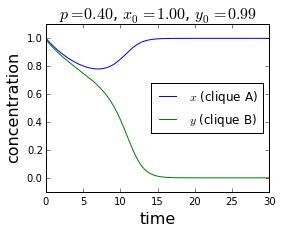} \\
\includegraphics[scale=0.5]{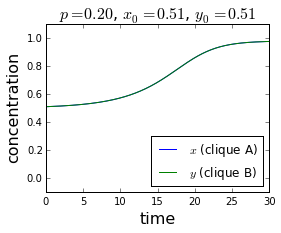} \  \includegraphics[scale=0.5]{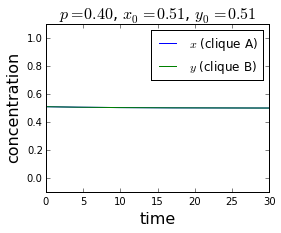} \\
\includegraphics[scale=0.5]{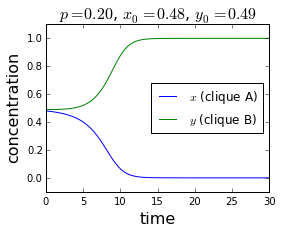} \  \includegraphics[scale=0.5]{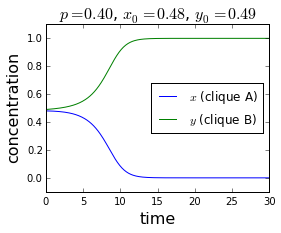} \\
\caption{Time evolution of the dynamical system given by Eq.~(\ref{eq:dynamical2}) for two different values of the parameter $p$: $0.2$ (left column) and $0.4$ (right column). The initial concentrations of spinsons in state $+1$ in clique $A$ (i.e. $x_0$) and clique $B$ ($y_0$) are given in the titles of the plots. The size of the influence group $q$ was set to 3 in all calculations. \label{fig:evolution}}
\end{figure}

The time evolution of our dynamical system for two different values of $p$ is shown in Figure~\ref{fig:evolution}.
The set~(\ref{eq:dynamical2}) of ordinary differential equations was solved numerically in Python by making use of the \texttt{odeint} function from the SciPy package~\cite{SCI:16}.

As already known from Figs.~\ref{fig:quiver} and~\ref{fig:bif}, at $p=0.2$ our system  may end up either in the polarized state (i.e. $P_1$ or $P_2$)
or in the consensus one ($C_1$ or $C_2$) depending on the initial conditions for the concentrations of $+1$ spinsons. We see that the results shown in the left column of Figure~\ref{fig:evolution} are in line with these findings. If for instance the starting point is the total positive consensus in clique $A$ ($x_0 = 1.00$) and almost total consensus in clique $B$ ($y_0=0.99$), then the system ends up in state $C_2 = (0.981,0.981)$ representing consensus in the entire system (top left plot in Figure~\ref{fig:evolution}). In this case the anticonformistic links between cliques, the number of which is represented by $p$, lead only to a tiny decrease of the initial concentrations of $+1$ agents in both cliques. If the initial conditions are close to the repeller $R_1$ and symmetric (i.e. we start from a point on the diagonal in the state plane, \color{black}which means that there is (almost) no ordering in each clique\color{black}), then again the system reaches the consensus state (middle left plot). However, a small  deviation from the diagonal pushes the system towards the polarized state (bottom left).

As we see, the behaviour for $p=0.4$ is different. The system usually ends up in the polarized state (top and middle right plots in Figure~\ref{fig:evolution}). The only exception are initial conditions along the diagonal in the state plane $(x,y)$. Since $R_1$ is now not a repeller, but a hyperbolic fixed point, the system is pushed towards it in this case (bottom right plot). Again, this is in agreement with the direction field shown in Figure~\ref{fig:quiver}.

\subsection{Basins of attraction}

Attractors of every dynamical system are surrounded by a basin of attraction representing the set of initial conditions in the state space whose orbits approach the attractor as time goes to infinity. In the previous paragraph we have seen already examples of initial conditions belonging to the basins of the polarization equilibria ($P_1$ or $P_2$) and the consensus ones ($C_1$ or $C_2$) (see Figure~\ref{fig:evolution}). Now, we would like to quantify the shapes of the basins of different attractors of our system. Of particular interest in a model with segmentation and negative ties are the basins of the consensus states, since in such a setting one intuitively expects polarization as the natural asymptotic state.

\begin{figure}[H]
\centering
\includegraphics[scale=0.5]{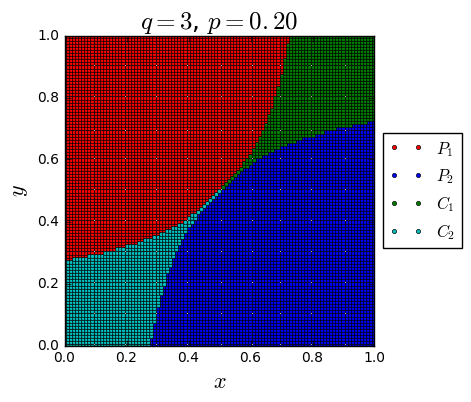} \  \includegraphics[scale=0.5]{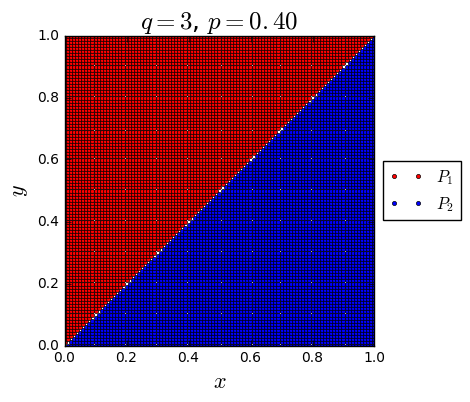} \\
\caption{Basins of attraction at two different values of $p$, $0.2$ (left plot) and $0.4$ (right plot). Each point in the state plane corresponds to an initial condition. Its color indicates the attractor, which a solution is converging to. The coordinates of the attractors are defined in Eqs.~(\ref{eq:fixed}) and~(\ref{eq:fixed2}). Boundaries between the basins are formed by stable manifolds of the hyperbolic points $U_1$--$U_4$ (left plot) or of the point $R_1$ (right plot).\label{fig:basins}}

\end{figure}

Since it is not possible to calculate the shapes analytically, we will resort to numerical methods again. For that purpose we create first a grid in the state plane $(x,y)$ with both coordinates varying from $0.0$ to $1.0$ with step $0.01$. The points on the grid represent different initial conditions uniformly distributed in the whole state plane. Then we solve the dynamical system~(\ref{eq:dynamical2}) for each grid point and check what attractor the solution is converging to at long evolution times. Results for $p=0.2$ and $p=0.4$ are presented in Figure~\ref{fig:basins}. As often in this paper, the  parameter $q$ was set to 3 in the whole procedure. 

For $p=0.2$ (left plot in Figure~\ref{fig:basins}) the whole state plane is divided into the basins of four attractors: $P_1$, $P_2$, $C_1$ and $C_2$. Their coordinates are defined in Eq.~(\ref{eq:fixed}). Surprisingly the basins of both positive ($C_2$) and negative ($C_1$) consensus are relatively large. Thus, if the number of negative connections between the cliques is small, the consensus may still by reached in a double-clique network for a range of initial conditions. It is worth to mention that the boundaries between the basins observed in the plot are formed by the stable manifolds of non-symmetric hyperbolic points $U_1$--$U_4$.  

The picture at $p=0.4$ (right plot in Figure~\ref{fig:basins}) is simpler. Consensus is no longer possible and the whole state plane is divided into the basins of two polarization points $P_1$ and $P_2$ (see Eq.~(\ref{eq:fixed2}) for their coordinates). The boundary between them corresponds to the stable manifold of $R_1$, which is now hyperbolic.

\subsection{Correlation between cliques}

All results presented up to this point indicate that some sort of a competition between conformity and anticonformity may be responsible for substantial changes in the dynamics of the model. To elaborate on that issue we will look at the product of the final states of the cliques,
\begin{equation}
c_A^\infty c_B^\infty = x^\infty y^\infty,
\end{equation}
as a function of $p$ at different values of $q$. We will call this quantity a correlation between the cliques. To allow for comparisons with the results presented in~\cite{SIE:SZW:WER:16} we will focus our attention on the total positive consensus $x_0,y_0=1.0$ as the initial condition. It is referred to as the Scenario I in~\cite{SIE:SZW:WER:16} and corresponds to the following situation: two cliques with a natural tendency to disagree with each other evolve at first independently. They get in touch by chance and establish some cross-links to the other group once they both reached consensus on a given issue.

\begin{figure}[H]
\centering
\includegraphics[scale=0.5]{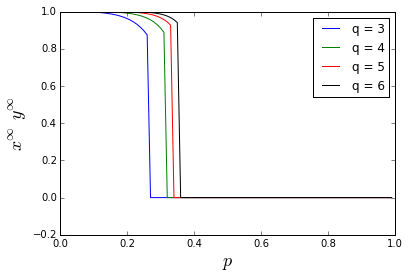}
\caption{Correlation $x^\infty y^\infty$ between final states of the cliques as a function of $p$ for different values of $q$. Below the bifurcation (critical) value of $p$ consensus between cliques is possible. With increasing $p$ the consensus is slightly diminished because of increasing role of the negative ties between the cliques. Above the bifurcation point only polarization is possible in the system. \label{fig:corr}}
\end{figure}

Results for the correlation between the cliques are presented in Figure~\ref{fig:corr}.
For values of $p$ smaller than a critical value ($\simeq 0.267$ for $q = 3$) both cliques always end up in positive consensus. In other words, in
this regime the intra-clique conformity wins with the inter-clique anticonformity and both communities are able to maintain their initial positive consensus. If the value of $p$ is larger than the critical one, the anticonformity induced effects take over and the whole system ends up in a polarized state. Moreover, the critical point shifts with increasing $q$ towards higher values of $p$ (see Table~\ref{tab:critical} for more details) . Thus, the bigger the influence group, the more cross-links are needed to polarize the society.
\begin{table}[H]
\centering
\begin{tabular}{c|cccc}
\hline \hline
$q$ & 3 & 4 & 5 & 6 \\ 
$p^*$ & 0.267 & 0.311 & 0.339 & 0.359 \\ 
\hline \hline
\end{tabular}
\caption{The values of $p$ at the bifurcation point for different values of $q$.\label{tab:critical}} 
\end{table}

Let us compare the above results with the model presented in~\cite{SIE:SZW:WER:16}. For that purpose we have to convert the concentrations of agents in state $+1$ in each clique into average opinions according to Eq.~(\ref{eq:m_c}) and to transform the probability $p$ of being connected to other clique into the number of cross-links $L$. We obtain the transformation formula immediately from Eq.~(\ref{eq:p}):
\begin{equation}
L=\frac{p}{1-p}
\end{equation}

\begin{figure}[H]
\centering
\includegraphics[scale=0.5]{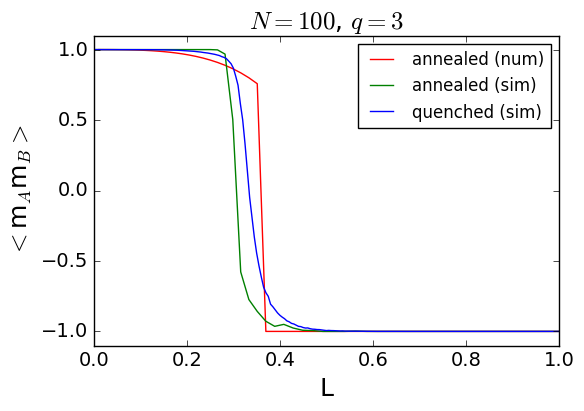} \ \includegraphics[scale=0.5]{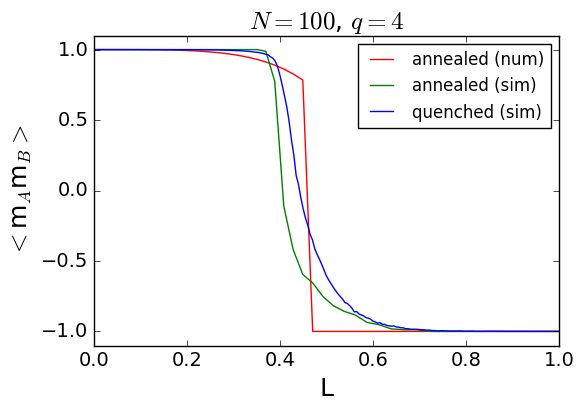}\caption{Correlation between cliques as a function of the fraction of cross-links $L$ for $q=3$ (left plot) and $q=4$ (right plot). Simulation results of the quenched disorder model (label `quenched (sim)' in the plots) taken from~\cite{SIE:SZW:WER:16} are compared with the numerical solution (`annealed (num)') of the asymtotic dynamical system given by Eq.~(\ref{eq:dynamical2}) as well as with the simulation of the annealed model as a birth-death process (`annealed (sim)'). Both simulations were performed for a finite size system ($N=100$ agents in each clique). Despite the differences between the models the agreement between them is quite good. In particular, for both values of $q$ the transition from consensus to polarization sets in at similar values of $L$.\label{fig:comparison}}
\end{figure}

The comparisons for two different values of $q$ are shown in Figure~\ref{fig:comparison}.
The agreement of the results is quite good, despite the differences between the models and the fact that the numerical solution for the annealed model was obtained for an infinite system, whereas the simulations were performed for only $N=100$ agents in each clique. Most notably, the transition from consensus to polartization sets in at similar values of $L$ in both models.

\begin{figure}[H]
\centering
\includegraphics[scale=0.5]{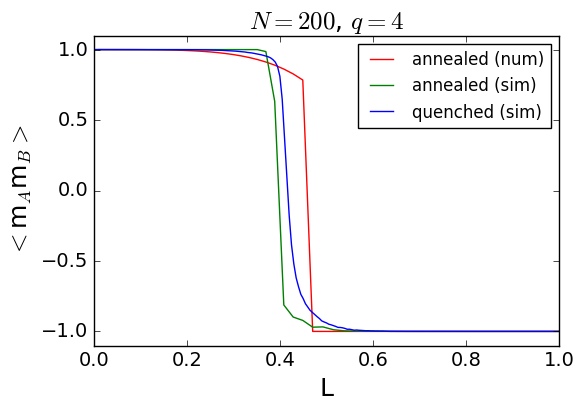} \ \includegraphics[scale=0.5]{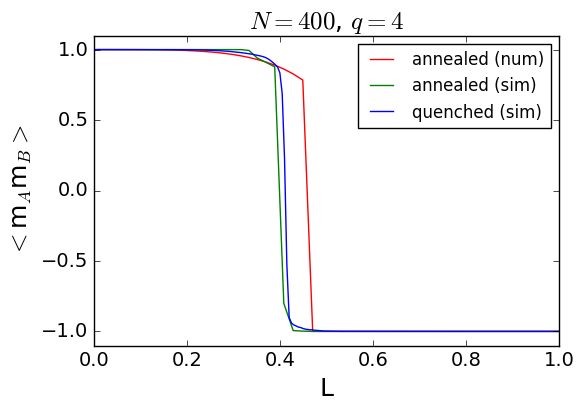}\\
\caption{Correlation between cliques as a function of the fraction of cross-links $L$ for $q=4$ and two different system sizes: $N=200$(left plot) and $N=400$ (right plot). See caption of Figure~\ref{fig:comparison} for explanation of the labels. The differences between simulation results for both models decrease with increasing system size.\label{fig:comparison2}}
\end{figure}

Some of the discrepancies between the numerical results for the annealed model and the simulation results for the quenched disorder one are due to the finite system size of the quenched disorder model and the stochastic nature of its simulation. Indeed, the agreement between models is even better, if we leave the numerical solution out of consideration for a moment and concentrate only on the simulation results in both cases.  Moreover, as it follows from Figure~\ref{fig:comparison2}, the differences between the models decrease with increasing system size. Thus, for $N\rightarrow\infty$ both variants of the model will probably converge to each other.  
%

\section{Discussion}

Due to the computational complexity of Monte Carlo simulation approach to agent-based models we were able to investigate only few distinct initial conditions in~\cite{SIE:SZW:WER:16}. In this paper, thank to a slight modification of the model, we could analyse it mathematically and therefore explore the whole state plane. 

Again, our results indicate that the \color{black}interplay \color{black} between conformity and anticonformity may lead to a polarized state of the system. We now have however much better understanding of the conditions necessary to arrive at consensus and we have determined regimes in which polarization takes over. Thus the present results complete the analysis started in~\cite{SIE:SZW:WER:16}. The most important message of the study is that consensus between two antagonistic communities is possible only if they are loosely connected with each other. 

It should be noted that similar results have been achieved by Shin and Lorenz~\cite{SHI:LOR:10} within the information accumulation system model of continuous opinion dynamics. The authors found that the convergence of two internally highly connected communities with a comparably low number of cross-links to the same opinion is less possible the more overall interaction between agents takes place.

\color{black}
Explosive growth in Internet mediated communication facilitates the exchange of opinions between people, both passively and actively~\cite{WIS:10}. Within the language of our model it means that modern communication tools like Facebook and Twitter increase the number of links between people in general and between members of different minded cliques in particular. Our results imply that if the fraction of cross links (i.e. the value of $p$) between such cliques exceeds some critical value, then the polarized state is the only attractor of the system. In real social systems the situation is of course more complicated, because the cliques interact not only with each other, but with other actors as well. However, our results may indicate one of the possible mechanisms for the omnipresence of polarization in social media. Paradoxically, the often criticized ``filter bubbles'' on Facebook or Google~\cite{BOZ:13}, which separate users from information (people) that disagrees with their viewpoints may help to weaken the problem with bi-polarization and to maintain overall consensus, because they reduce the fraction of cross links between cliques (the discussion of negative effects of ``filter bubbles'' on the society is beyond the scope of this work). 
\color{black}

As far as potential extensions to our model are concerned, the suggestions given in our previous paper still hold. For instance, it could be very informative to check how robust the model is to the introduction of noise, because it is already known that including noise to models of opinion dynamics may significantly change  their predictions~\cite{KLE:EGU:TOR:MIG:03}. Another interesting aspect worth to address in future studies is the casual connection between the network segmentation and the polarization.

\vspace{6pt} 

\supplementary{A Jupyter notebook with Python code used for the numerical analysis of the dynamical system~(\ref{eq:dynamical2}) may be found at \url{https://doi.org/10.5281/zenodo.167817}}

\acknowledgments{This work was partially supported by funds from the Polish National Science Centre (NCN) through grant no. 2013/11/B/HS4/01061.}

\authorcontributions{ T.K. and J.S. conceived and designed the experiments; T.K. performed the formal analysis of the annealed model; T.W. implemented the models and run  computer simulations; J.S. analyzed the annealed model numerically; T.W. and J.S. compiled and visualized the results; J.S. wrote the paper.}

\conflictsofinterest{The authors declare no conflict of interest.} 


\appendixtitles{yes} 
\appendixsections{multiple} 
\appendix

\section{Critical values of $p$ in case $q=2$}
\label{q2}

From the analysis presented in the main text (see Sec.~\ref{phase}) it follows that the system undergoes two phase transitions. At the first critical point four hyperbolic nonsymmetric fixed points disappear and the nontrivial symmetric equilibria become hyperbolic.  At the second critical point these symmetric points disappear as well. 

Our goal is to find both critical values of $p$ analytically in the special case $q=2$. We will proceed by first computing the coordinates of the
nontrivial symmetric fixed points as  functions of $p$. Then we find coordinates of a symmetric point, at which the largest eigenvalue of the Jacobian is equal to zero. Combining these two results will give us the critical value of the first transition. The second phase transition occurs at a value of $p$, for which both symmetric points disappear (i.e. they become complex).

For computing the nontrivial symmetric fixed points we put $x=y$ and $q=2$ into Eq.~(\ref{eq:dynamical2}). Setting $x^\prime$ to zero yields
\begin{equation}
\left( 1-x\right) \left( \left( 1-p\right) x+p\left( 1-x\right) \right)
^{2}-x\left( \left( 1-p\right) \left( 1-x\right) +px\right) ^{2}=0
\end{equation}
Factoring out and simplifying the above equation gives 
\begin{equation}
4px-x+p^{2}+3x^{2}-2x^{3}-12px^{2}-6p^{2}x+8p
x^{3}+12p^{2}x^{2}-8p^{2}x^{3}=0,
\end{equation}
which may be written as
\begin{equation}
\left( 1-2x\right) \left(
4px-x+p^{2}+x^{2}-4px^{2}-4p^{2}x+4p^{2}x^{2}\right) =0.
\end{equation}
Thus
\begin{equation}
4px-x+p^{2}+x^{2}-4px^{2}-4p^{2}x+4p^{2}x^{2} =  0, 
\end{equation}
and after some calculations we arrive at
\begin{equation}
x^{2}-x+\frac{p^{2}}{\left( 2p-1\right) ^{2}} = 0.
\label{xeqn}
\end{equation}
Solving the last equation gives 
\begin{eqnarray}
x_{1} & = & \frac{1}{2}-\frac{1}{2}\sqrt{1-\frac{4p^{2}}{\left( 2p-1\right) ^{2}}},\\
x_{2} & = & \frac{1}{2}+\frac{1}{2}\sqrt{1-\frac{4p^{2}}{\left( 2p-1\right) ^{2}}}.
\end{eqnarray}
Note that for $p > \frac{1}{4}$ both solutions are complex (i.e. the symmetric points disappear). Hence the second transition occurs at
\begin{equation}
p_2^\star = \frac{1}{4}.
\end{equation}

We proceed by computing the Jacobian at a symmetric point $\left( x,x\right) $. Due to the symmetry of the dynamical system~(\ref{eq:dynamical2}) the Jacobian at such a point is of the form
\begin{equation}
\left|
\begin{array}{ll}
a(x) & b(x) \\
b(x) & a(x)
\end{array}
\right|.
\label{jac}
\end{equation}
Hence the eigenvalues are (we omit the explicit dependency on $x$ for the sake of simplicity)
\begin{eqnarray}
\lambda_1 & = & a + b,\\
\lambda_2 & = & a - b.
\end{eqnarray} 
After some straightforward but lengthy computation we get
\begin{eqnarray}
a & = & \left( 1-2p\right) \left( 2x^{2}\left( 4p-3\right) -2x\left( 4p-3\right) -1+2p\right),\\
b & = & -2p\left( 2x\left( 1-x\right) \left( 1-2p\right) +p\right)
\end{eqnarray}
Since $b$ is negative for $p < \frac{1}{2}$,  the largest eigenvalue is always $a-b$. Setting it to zero yields
\begin{equation}
x^{2}-x+\frac{1+2p^{2}-4p}{8p^{2}-16p+6}=0
\end{equation}
Combining the last expression with Eq.~(\ref{xeqn}) gives
\begin{equation}
\frac{1+2p^{2}-4p}{8p^{2}-16p+6}=\frac{p^{2}}{\left( 2p-1\right) ^{2}},
\end{equation}
which is equivalent to
\begin{equation}
16p^{2}-8p-8p^{3}+1=0.
\end{equation}
We have to solve the above equation in order to get the critical value for the first transition. The relevant solution is
\begin{equation}
p_1^\star = \frac{3}{4}-\frac{1}{4}\sqrt{5}\simeq \allowbreak 0.190\,98.
\end{equation}

\section{Second critical value of $p$ for general $q$}
\label{qall}

The critical value of $p$ for the second phase transition may be calculated analytically in case of a general $q$. At that critical point the equilibrium $
R_1 = (1/2,1/2)$ changes its character from a repeller to a hyperbolic one. Hence, we have to evaluate the Jacobian at $R_1$ and look for a value of $p$, for which the smallest eigenvalue becomes zero.

The Jacobian will again have the form given by Eq.~(\ref{jac}). Taking the derivatives of the right-hand side of Eq.~(\ref{eq:dynamical2}) with respect to $x$ and $y$ and setting $x=y=1/2$ in the resulting expressions give
\begin{eqnarray}
a & = & 2q\left( \frac{1}{2}\right) ^{q}-2\left( \frac{1}{2}\right) ^{q}-2pq\left(\frac{1}{2}\right)^{q},\\
b & = & -pq\left( \frac{1}{2}\right) ^{q-1}.
\end{eqnarray}
The smallest eigenvalue of the Jacobi matrix is given by 
\begin{equation}
a+b=2q\left( \frac{1}{2}\right) ^{q}-2\left( \frac{1}{2}\right)
^{q}-2pq\left( \frac{1}{2}\right) ^{q}-pq\left( \frac{1}{2}\right) ^{q-1}.
\end{equation}
Equating it to zero yields the critical value of $p$,
\begin{equation}
p_2^\star = \frac{2\left( \frac{1}{2}\right) ^{q}-2q\left( \frac{1}{2}\right) ^{q}}{
-2q\left( \frac{1}{2}\right) ^{q}-q\left( \frac{1}{2}\right) ^{q-1}}
\end{equation}
For $q=2$ we obtain
\begin{equation}
p_2^\star = \frac{1}{4},
\end{equation}
which agrees with the result obtained in the previous section.

\section{Supplementary materials}

A Jupyter notebook with Python code used for the numerical analysis of the dynamical system~(\ref{eq:dynamical2}) may be found in~\citep{KRU:SZW:WER:16}.




\externalbibliography{yes}
\bibliography{polarization.bib}


\end{document}